\documentclass{birkmult}

\newcommand{\SetFigFont}[3]{}
\newcommand{\Og}{{\mathcal{O}}}

\newtheorem{Def}{Definition}[section]
\newtheorem{Thm}[Def]{Theorem}

\newtheorem{Conjecture}[Def]{Conjecture}

\newcommand{\spc}{\;\;\;\;\;\;\;\;\;\;}

\newcommand{\bra}{\mbox{$< \!\!$ \nolinebreak}}
\newcommand{\ket}{\mbox{\nolinebreak $>$}}
\newcommand{\C}{\mathbb{C}}
\newcommand{\R}{\mathbb{R}}
\newcommand{\1}{\mbox{\rm 1 \hspace{-1.05 em} 1}}

\newcommand{\Pdd}{\mbox{$\partial$ \hspace{-1.2 em} $/$}}
\newcommand{\slsh}{\mbox{ \hspace{-1.1 em} $/$}}
\newcommand{\Tr}{\mbox{\rm{Tr}\/}}
\newcommand{\beq}{\begin{equation}}
\newcommand{\eeq}{\end{equation}}

\newcommand{\M}{{\mathcal{M}}}
\newcommand{\tM}{\tilde{\mathcal{M}}}
\newcommand{\hM}{\hat{\mathcal{M}}}

\newcommand{\Lat}{{\mathfrak{L}}}
\newcommand{\m}{{\mathfrak{m}}}

\begin{document}
%
\firstpage{1}
%
%
%
%
%
%
%
%
\title{From Discrete Space-Time to Minkowski Space:
Basic Mechanisms, Methods and Perspectives}
\author{Felix Finster}

\address{%
NWF I -- Mathematik \\
Universit\"at Regensburg \\
D-93040 Regensburg, Germany}

\email{Felix.Finster@mathematik.uni-regensburg.de}
\thanks{I would like to thank the organizers of the workshop ``Recent Developments in Quantum Field Theory'' (Leipzig, July 2007) for helpful discussions and encouragement.
I thank Joel Smoller and the referee for helpful comments on the manuscript.}

\markboth{Felix Finster}{From Discrete Space-Time to Minkowski Space}

\maketitle
\begin{abstract}
This survey article reviews recent results on fermion system in discrete space-time
and corresponding systems in Minkowski space.
After a basic introduction to the discrete setting, we explain
a mechanism of spontaneous symmetry breaking which leads to the emergence
of a discrete causal structure.
As methods to study the transition between discrete space-time and Minkowski
space, we describe a lattice model for a static and isotropic space-time,
outline the analysis of regularization tails of vacuum Dirac sea configurations, and introduce
a Lorentz invariant action for the masses of the Dirac seas.
We mention the method of the continuum limit, which allows to analyze
interacting systems. Open problems are discussed.
\end{abstract}
\tableofcontents

\section{Introduction} \label{sec0}
\markboth{Felix Finster}{From Discrete Space-Time to Minkowski Space}
It is generally believed that the concept of a space-time continuum
(like Minkowski space or a Lorentzian manifold) should be modified
for distances as small as the Planck length. The principle of the
fermionic projector~\cite{PFP} proposes a new model of space-time,
which should be valid down to the Planck scale.
This model is introduced as a system of quantum mechanical wave
functions defined on a finite number of space-time points and is referred
to as a {\em{fermion system in discrete space-time}}.
The interaction is described via a variational principle where we minimize an action defined
for the ensemble of wave functions.
A-priori, there are no relations between the space-time points; in particular, there is
no nearest-neighbor relation and no notion of causality. The idea is that these 
additional structures should be generated spontaneously.
More precisely, in order to minimize the action, the wave functions form specific
configurations; this can be visualized as a ``self-organization'' of the particles.
As a consequence of this self-organization, the wave functions induce non-trivial relations
between the space-time points. We thus obtain additional structures in space-time,
and it is conjectured that, in a suitable limit where the number of particles and space-time points
tends to infinity, these structures should give rise to the local and causal structure
of Minkowski space. In this limit, the configuration of the wave functions should
go over to a Dirac sea structure.

This conjecture has not yet been proved, but recent results give a detailed
picture of the connection between discrete space-time and Minkowski space.
Also, mathematical methods were developed to shed light on particular aspects of the problem.
In this survey article we report on the present status, explain basic mechanisms
and outline the analytical methods used so far. The presentation is self-contained and non-technical.
The paper concludes with a discussion of open problems.

\section{Fermion Systems in Discrete Space-Time} \label{sec1}
\markboth{Felix Finster}{From Discrete Space-Time to Minkowski Space}
We begin with the basic definitions in the discrete setting (for more details see~\cite{F1}).
Let~$(H, \bra .|. \ket)$ be a finite-dimensional complex inner product space.
Thus~$\bra .|. \ket$ is linear in its second and anti-linear in its first argument, and it is symmetric,
\[ \overline{\bra \Psi \:|\: \Phi \ket} \;=\; \bra \Phi \:|\: \Psi \ket \quad
\spc {\mbox{for all~$\Psi,\Phi \in H$}} \,, \]
and non-degenerate,
\[ \bra \Psi \:|\: \Phi \ket \;=\; 0 \;\;\; {\mbox{for all $\Phi \in H$}}
 \quad \Longrightarrow \quad
\Psi \;=\; 0 \:. \]
In contrast to a scalar product, $\bra .|. \ket$ need {\em{not}} be positive.

A {\em{projector}}~$A$ in~$H$ is defined just as in Hilbert spaces as a linear
operator which is idempotent and self-adjoint,
\[ A^2 = A \spc {\mbox{and}} \spc \bra A\Psi \:|\: \Phi \ket = \bra \Psi \:|\: A\Phi \ket \quad
{\mbox{for all $\Psi, \Phi \in H$}}\:. \]
Let~$M$ be a finite set. To every point~$x \in M$ we associate a projector
$E_x$. We assume that these projectors are orthogonal and
complete in the sense that
\beq \label{oc}
E_x\:E_y \;=\; \delta_{xy}\:E_x \spc {\mbox{and}} \spc
\sum_{x \in M} E_x \;=\; \1\:.
\eeq
Furthermore, we assume that the images~$E_x(H) \subset H$ of these
projectors are non-de\-ge\-ne\-rate subspaces of~$H$, which
all have the same signature~$(n,n)$.
We refer to~$n$ as the {\em{spin dimension}}.
The points~$x \in M$ are
called {\em{discrete space-time points}}, and the corresponding
projectors~$E_x$ are the {\em{space-time projectors}}. The
structure~$(H, \bra .|. \ket, (E_x)_{x \in M})$ is
called {\em{discrete space-time}}.

We next introduce the so-called {\em{fermionic projector}} $P$ as
a projector in~$H$ whose image~$P(H) \subset H$ is
{\em{negative definite}}.
The vectors in the image of~$P$ have the interpretation as the
quantum states of the particles of our system. Thus the rank of~$P$
gives the {\em{number of particles}} $f := \dim P(H)$.
The name ``fermionic projector'' is motivated from the correspondence to
Minkowski space, where our particles should go over to Dirac particles,
being fermions (see Section~\ref{sec5} below).
We call the obtained structure~$(H, \bra .|. \ket, (E_x)_{x \in M}, P)$
a {\em{fermion system in discrete space-time}}.
Note that our definitions involve only
three integer parameters: the spin dimension~$n$, the number
of space-time points~$m$, and the number of particles~$f$.

The above definitions can be understood
as a mathematical reduction to some of the structures present in relativistic
quantum mechanics, in such a way that the
{\em{Pauli Exclusion Principle}}, a {\em{local gauge principle}} and
the {\em{equivalence principle}} are respected (for details see~\cite[Chapter~3]{PFP}).
More precisely, describing the many-particle system by a projector~$P$,
every vector $\Psi \in H$ either lies in the image of $P$ or it does not. In this way, the
fermionic projector encodes for every state the occupation numbers $1$ and $0$, respectively,
but it is impossible to describe higher occupation numbers.
More technically, choosing a basis~$\Psi_1, \ldots \Psi_f$ of~$P(H)$, we can
form the anti-symmetric many-particle wave function
\[ \Psi \;=\; \Psi_1 \wedge \cdots \wedge \Psi_f \:. \]
Due to the anti-symmetrization, this definition of~$\Psi$ is (up to a phase)
independent of the choice of the basis $\Psi_1,\ldots, \Psi_f$.
In this way, we can associate to every fermionic projector a fermionic
many-particle wave function, which clearly respects the Pauli Exclusion principle.
To reveal the local gauge principle, we consider unitary operators~$U$
(i.e.\ operators which for all~$\Psi, \Phi \in H$ satisfy
the relation $\bra U \Psi | U \Phi \ket = \bra \Psi | \Phi \ket$) which 
do not change the space-time projectors,
\beq \label{gauge1}
E_x \;=\; U E_x U^{-1} \spc {\mbox{for all $x \in M$}}\:.
\eeq
We transform the fermionic projector according to
\beq \label{gauge2}
P \;\to\; U P U^{-1}\:.
\eeq
Such transformations lead to physically equivalent fermion systems.
The conditions~(\ref{gauge1}) mean that~$U$ maps every subspace~$E_x(H)$ onto itself. In other words, $U$ acts ``locally'' on the subspaces associated to the
individual space-time points. The transformations~(\ref{gauge1}, \ref{gauge2})
can be identified with local gauge transformations in physics
(for details see~\cite[\S3.1]{PFP}).
The equivalence principle is built into our framework in a very general form
by the fact that our definitions do not distinguish an ordering between the space-time
points. Thus our definitions are symmetric under permutations of the space-time points,
generalizing the diffeomorphism invariance in general relativity.

Obviously, important physical principles are missing in our framework.
In particular, our definitions involve {\em{no locality}} and {\em{no causality}},
and not even relations like the nearest-neighbor relations on a lattice.
The idea is that these additional structures, which are of course essential for
the formulation of physics, should emerge as a consequence of a spontaneous
symmetry breaking and a self-organization of the particles as described by
a variational principle. Before explaining in more detail how this is supposed
to work (Section~\ref{sec5}), we first introduce the variational principle (Section~\ref{sec2}),
explain the mechanism of spontaneous symmetry breaking (Section~\ref{sec3}),
and discuss the emergence of a discrete causal structure (Section~\ref{sec4}).

\section{A Variational Principle} \label{sec2}
In order to introduce an interaction of the particles, we now set up a variational principle. 
For any~$u \in H$, we refer to the projection~$E_x u \in E_x(H)$ as the
{\em{localization}} of~$u$ at~$x$. We also use the short notation~$u(x)=E_x u$ and
sometimes call~$u(x)$ the {\em{wave function}} corresponding to the vector~$u$.
Furthermore, we introduce the short notation
\beq \label{notation}
P(x,y) \;=\; E_x\,P\,E_y\:,\spc x,y \in M \:.
\eeq
This operator product maps~$E_y(H) \subset H$ to~$E_x(H)$, and it is often
useful to regard it as a mapping only between these subspaces,
\[ P(x,y)\;:\; E_y(H) \: \rightarrow\: E_x(H)\:. \]
Using the properties of the space-time projectors~(\ref{oc}), we find
\[ (Pu)(x) \;=\; E_x\: Pu \;=\; \sum_{y \in M} E_x\,P\,E_y\:u
\;=\; \sum_{y \in M} (E_x\,P\,E_y)\:(E_y\,u) \:, \]
and thus
\beq \label{diskernel}
(Pu)(x) \;=\; \sum_{y \in M} P(x,y)\: u(y)\:.
\eeq
This relation resembles the representation of an operator with an integral kernel,
and thus we refer to~$P(x,y)$ as the {\em{discrete kernel}} of the fermionic projector.
Next we introduce the {\em{closed chain}} $A_{xy}$ as the product
\beq \label{cc}
A_{xy} \;:=\; P(x,y)\: P(y,x) \;=\; E_x \:P\: E_y \:P\: E_x \:;
\eeq
it maps~$E_x(H)$ to itself.
Let~$\lambda_1,\ldots,\lambda_{2n}$ be the roots of the characteristic polynomial
of~$A_{xy}$, counted with multiplicities. We define the {\em{spectral weight}}~$|A_{xy}|$
by
\[ |A_{xy}| \;=\; \sum_{j=1}^{2n} |\lambda_j|\:. \]
Similarly, one can take the spectral weight of powers of~$A_{xy}$, and by summing
over the space-time points we get positive numbers depending only on the
form of the fermionic projector relative to the space-time projectors.
Our variational principle is to
\beq \label{vary}
{\mbox{minimize}} \quad \sum_{x,y \in M} |A_{xy}^2|
\eeq
by considering variations of the fermionic projector which satisfy for a given real
parameter~$\kappa$ the constraint
\beq \label{constraint}
\sum_{x,y \in M} |A_{xy}|^2 = \kappa \:.
\eeq
In the variation we also keep the number of particles~$f$ as well as
discrete space-time fixed. Clearly, we need to choose~$\kappa$ such that there is at least one fermionic projector which satisfies~(\ref{constraint}).
It is easy to verify that~(\ref{vary}) and~(\ref{constraint})
are invariant under the transformations~(\ref{gauge1}, \ref{gauge2}),
and thus our variational principle is gauge invariant.

The above variational principle was first introduced in~\cite{PFP}. In~\cite{F1}
it is analyzed mathematically, and it is shown in particular that
minimizers exist:
\begin{Thm} \label{thmn1}
The minimum of the variational principle~(\ref{vary}, \ref{constraint}) is attained.
\end{Thm}

Using the method of Lagrange multipliers, for every minimizer~$P$ there is a real
parameter~$\mu$ such that~$P$ is a stationary point of the {\em{action}}
\beq \label{Sdef}
{\mathcal{S}}_\mu[P] \;=\; \sum_{x,y \in M} {\mathcal{L}}_\mu[A_{xy}]
\eeq
with the {\em{Lagrangian}}
\beq \label{Ldef}
{\mathcal{L}}_\mu[A] \;=\; |A^2| - \mu\: |A|^2 \:.
\eeq
A useful method for constructing stationary points for a
given value of the Lagrange multiplier~$\mu$ is to minimize
the action~${\mathcal{S}}_\mu$ without the constraint~(\ref{constraint}).
This so-called {\em{auxiliary variational principle}} behaves differently
depending on the value of~$\mu$. If~$\mu < \frac{1}{2n}$, the action is
bounded from below, and it is proved in~\cite{F1} that minimizers exist.
In the case $\mu > \frac{1}{2n}$, on the other hand, the action is not bounded from below,
and thus there are clearly no minimizers. In the remaining so-called {\em{critical case}}
$\mu=\frac{1}{2n}$, partial existence results are given in~\cite{F1}, but the general
existence problem is still open. The critical case
is important for the physical applications. For simplicity, we omit the
subscript~$\mu=\frac{1}{2n}$ and also refer to the auxiliary variational principle
in the critical case as the {\em{critical variational principle}}.
Writing the critical Lagrangian as
\beq \label{Lcrit}
{\mathcal{L}}[A] \;=\;
\frac{1}{4n} \sum_{i,j=1}^{2n} \left( |\lambda_i| - |\lambda_j| \right)^2 ,
\eeq
we get a good intuitive understanding of the critical variational principle:
it tries to achieve that for every $x, y \in M$, all the roots of the characteristic polynomial of
the closed chain~$A_{xy}$ have the same absolute value.

We next derive the corresponding {\em{Euler-Lagrange equations}}
(for details see~\cite[\S3.5 and \S5.2]{PFP}).
Suppose that~$P$ is a critical point of the action~(\ref{Sdef}).
We consider a variation~$P(\tau)$ of projectors with~$P(0)=P$.
Denoting the gradient of the Lagrangian by~$\mathcal{M}$,
\beq \label{fvar}
{\mathcal{M}}_\mu[A]^\alpha_\beta \;:=\;
\frac{\partial {\mathcal{L}}_\mu[A]}{\partial A^\beta_\alpha}\:, \qquad 
{\mbox{with }} \alpha, \beta
\in \{1,\ldots, 2n\}\:,
\eeq
we can write the variation of the Lagrangian as a trace on~$E_x(H)$,
\[ \delta {\mathcal{L}}_\mu[A_{xy}] \;=\; \frac{d}{d \tau}  {\mathcal{L}}_\mu[A_{xy}(\tau)]
\Big|_{\tau=0} \;=\; \Tr \left( E_x\,{\mathcal{M}}_\mu[A_{xy}] \:\delta A_{xy} \right) . \]
Using the Leibniz rule
\[ \delta A_{xy} \;=\; \delta P(x,y) \:P(y,x) \:+\: P(x,y) \:\delta P(y,x) \]
together with the fact that the trace is cyclic, after summing over the space-time points
we find
\[ \sum_{x,y \in M} \delta {\mathcal{L}}_\mu[A_{xy}] \;=\; \sum_{x,y \in M} 
4 \: \Tr \left( E_x\, Q_\mu(x,y) \:\delta P(y,x) \right)\:, \]
where we set
\beq \label{Qxydef}
Q_\mu(x,y) \;=\; \frac{1}{4} \left( {\mathcal{M}}_\mu[A_{xy}]\:P(x,y)
\:+\: P(x,y) \:{\mathcal{M}}_\mu[A_{yx}] \right) \:.
\eeq
Thus the first variation of the action can be written as
\beq \label{dS1}
\delta {\mathcal{S}}_\mu[P] \;=\; 4 \,\Tr \left( Q_\mu\: \delta P \right) ,
\eeq
where~$Q_\mu$ is the operator in~$H$ with kernel~(\ref{Qxydef}).
This equation can be simplified using
that the operators~$P(\tau)$ are all projectors of fixed rank.
Namely, there is a family of unitary operators~$U(\tau)$ with~$U(\tau)=\1$ and
\[ P(\tau) \;=\; U(\tau)\, P\, U(\tau)^{-1}\:. \]
Hence~$\delta P = i[B, P]$, where we set~$B=-iU'(0)$.
Using this relation in~(\ref{dS1}) and again using that the trace is
cyclic, we find $\delta {\mathcal{S}}_\mu[P] = 4i \,\Tr \left( [P,Q_\mu]\: B \right)$.
Since~$B$ is an arbitrary self-adjoint operator, we conclude that
\beq \label{EL}
[P,Q_\mu] \;=\; 0\:.
\eeq
This commutator equation with~$Q_\mu$ given by~(\ref{Qxydef}) are the
Euler-Lagrange equations corresponding to our variational principle.

\section{A Mechanism of Spontaneous Symmetry Breaking} \label{sec3}
In the definition of fermion systems in discrete space-time, we did not distinguish
an ordering of the space-time points; all our
definitions are symmetric under permutations of the points of~$M$.
However, this does not necessarily mean that a given fermion system will have this permutation
symmetry. The simplest counterexample is to take a fermionic projector consisting of
one particle which is localized at the first space-time point, i.e.\ in bra/ket-notation
\begin{eqnarray*}
P &=& -|u \ket \bra u | \qquad{\mbox{with}}\qquad \bra u \,|\, u \ket = -1 \qquad {\mbox{and}} \\
E_1 u &=& u \:, \qquad E_x u \;=\; 0 \quad {\mbox{for all $x = 2, \ldots, m$}}\:.
\end{eqnarray*}
Then the fermionic projector distinguishes the first space-time point and thus breaks
the permutation symmetry.
In~\cite{F2} it is shown under general assumptions on the number of particles and
space-time points that, no matter how we choose
the fermionic wave functions, it is impossible to arrange that the
fermionic projector respects the permutation symmetry.
In other words, the fermionic projector necessarily breaks the permutation
symmetry of discrete space-time. We first specify the result and explain it afterwards.
The group of all permutations of the space-time points is the symmetric group,
denoted by~$S_m$.
\begin{Def} \label{defouter}
A subgroup~$\Og \subset S_m$
is called {\bf{outer symmetry group}} of the fermion system in discrete space-time if
for every~$\sigma \in \Og$ there is a unitary transformation~$U$ such that
\beq \label{USdef}
UPU^{-1} \;=\; P \qquad {\mbox{and}} \qquad
U E_x U^{-1} \;=\; E_{\sigma(x)} \quad
{\mbox{for all $x \in M$}}\:.
\eeq
\end{Def}

\begin{Thm} {\bf{(spontaneous breaking of the  permutation symmetry)}} \label{thmend}
Suppose that $(H, \bra .|. \ket, (E_x)_{x \in M}, P)$ is a
fermion system in discrete space-time of spin dimension~$n$. Assume that the number of
space-time points~$m$ is sufficiently large,
\beq \label{nmbound}
m \;>\; \left\{ \begin{array}{ccl}
3 && {\mbox{if~$n=1$}} \\[.2em]
\displaystyle \max \Big(2n+1,\: 4 \,[\log_2 n] + 6 \Big) &\quad& {\mbox{if~$n>1$}}
\end{array} \right.
\eeq
(where~$[x]$ is the Gau{\ss} bracket),
and that the number of particles~$f$ lies in the range
\beq \label{fbound}
n \;<\; f \;<\; m-1 \:.
\eeq
Then the fermion system cannot have the outer symmetry group~$\Og = S_m$.
\end{Thm}

For clarity we note that this theorem does not refer to the variational principle
of Section~\ref{sec2}. To explain the result, we now give an alternative proof
in the simplest situation where the theorem applies:
the case~$n=1$, $f=2$ and $m=4$.
For systems of two particles, the following construction from~\cite{DFS}
is very useful for visualizing the fermion system.
The image of~$P$ is a two-dimensional, negative definite subspace of~$H$.
Choosing an orthonormal basis $(u_1,u_2)$ (i.e.\ $\bra u_i | u_j \ket = -\delta_{ij}$),
the fermionic projector can be written in bra/ket-notation as
\beq \label{2braket}
P \;=\; -|u_1 \ket \bra u_1| - |u_2 \ket \bra u_2|\:.
\eeq
For any space-time point~$x \in M$ we introduce the
so-called {\em{local correlation matrix}} $F_x$ by
\beq \label{Fxdef}
(F_x)^i_j \;=\; -\bra u_i \,|\, E_x u_j \ket \:.
\eeq
The matrix~$F_x$ is Hermitian on the standard Euclidean $\C^2$. Thus we
can decompose it in the form
\beq \label{Fxd}
F_x \;=\; \frac{1}{2} \left( \rho_x \1 + \vec{v}_x \vec{\sigma} \right) ,
\eeq
where~$\vec{\sigma}=(\sigma^1, \sigma^2, \sigma^3)$ are the Pauli matrices.
We refer to the~$\vec{v}_x$ as the {\em{Pauli vectors}}.
The local correlation matrices are obviously invariant under unitary transformations in~$H$.
But they do depend on the arbitrariness in choosing the orthonormal
basis~$(u_1, u_2)$ of~$P(H)$. More precisely, the choice of the orthonormal basis
involves a $U(2)$-freedom and, according to the transformation of Pauli spinors in
non-relativistic quantum mechanics, this gives rise to orientation preserving rotations
of all Pauli vectors. Hence the local correlation matrices are unique up to
the transformations
\beq \label{Rot}
\vec{v}_x \;\longrightarrow\; R\, \vec{v}_x
\qquad {\mbox{with}} \qquad R \in SO(3)\:.
\eeq
Let us collect a few properties of the local correlation matrices. Summing over~$x$
and using the completeness relation~(\ref{oc}), we find that~$\sum_{x \in M} F_x =\1$ or, equivalently,
\beq \label{rhovcond}
\sum_{x \in M} \rho_x \;=\; 2 \qquad {\mbox{and}} \qquad \sum_{x \in M} \vec{v}_x
\;=\; \vec{0}\:.
\eeq
Furthermore, as the inner product in~(\ref{Fxdef}) has signature~$(1,1)$, the
matrix~$F_x$ can have at most one positive and at most one negative eigenvalue.
Expressed in terms of the decomposition~(\ref{Fxd}), this means that
\beq \label{rhovc2}
|\vec{v}_x| \;\geq\; \rho_x \qquad {\mbox{for all $x \in M$}}\:.
\eeq

\begin{figure}[t]
\begin{center}
 \includegraphics[width=4cm]{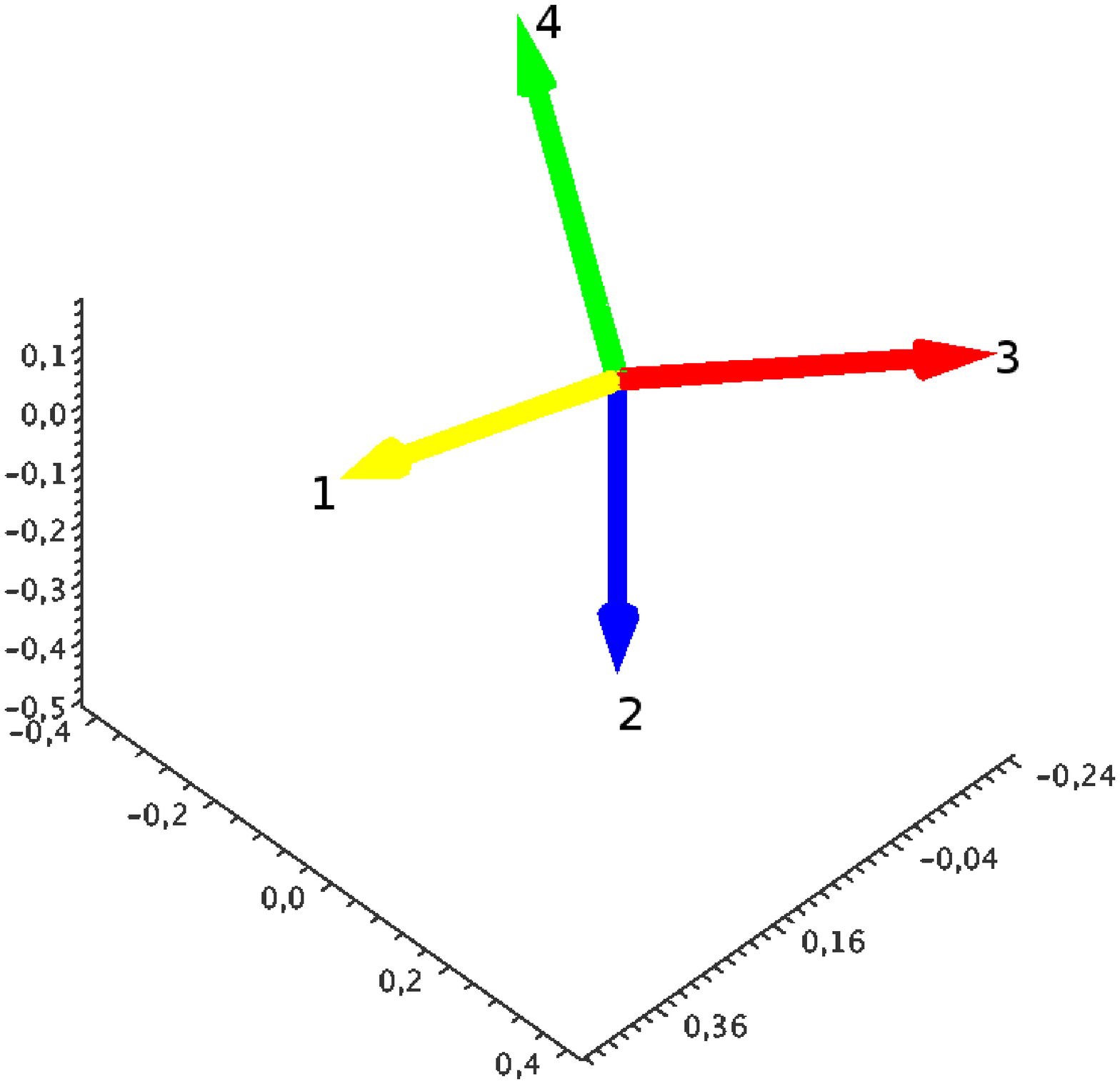} \hspace*{1cm}
 \includegraphics[width=4cm]{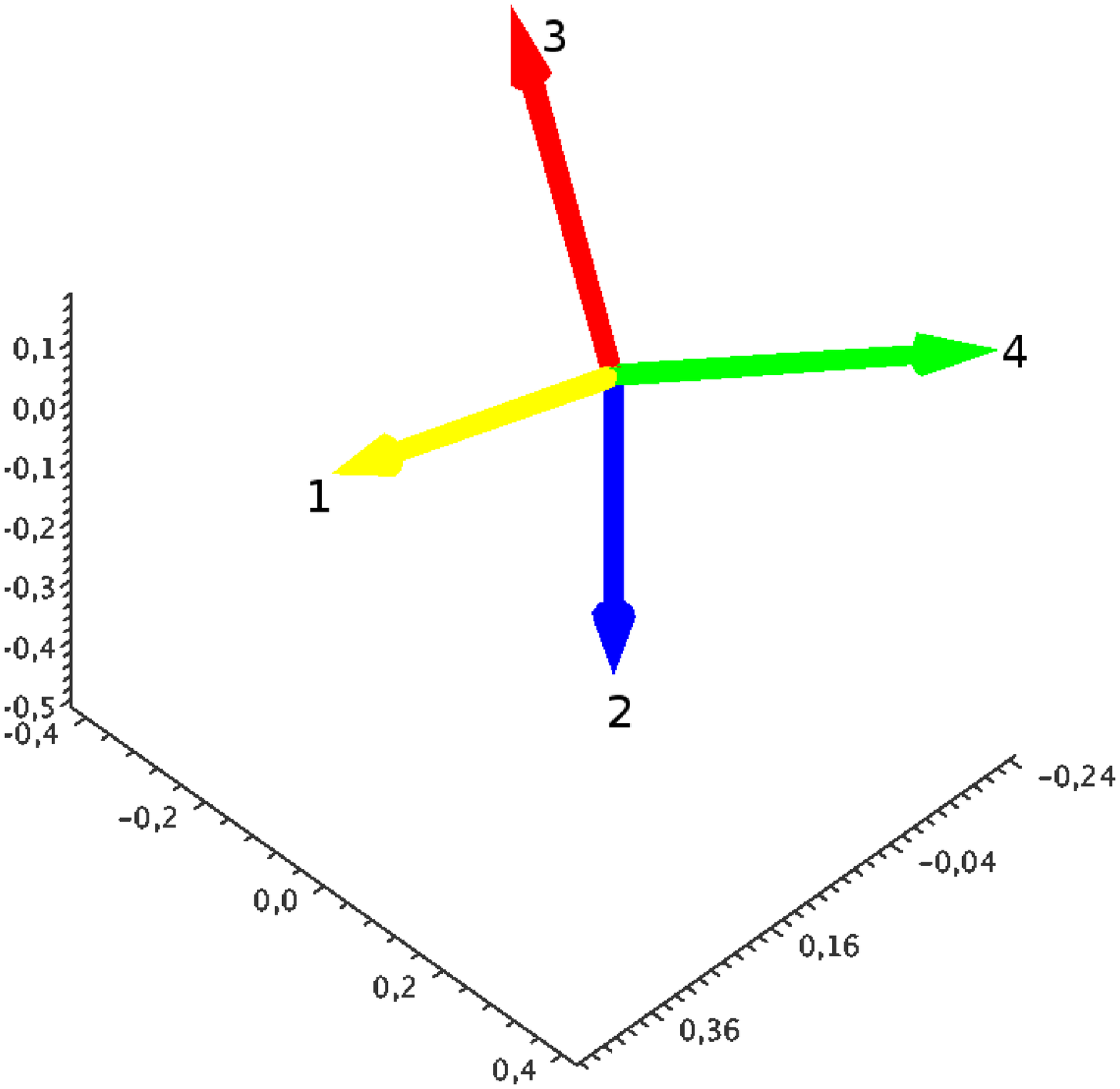}
 \caption{Tetrahedron configurations of the Pauli vectors} \label{fig1}
\end{center}
\end{figure}

Now assume that a fermion system with~$m=4$ space-time points is permutation symmetric.
Then the scalars~$\rho_x$ must all be equal. Using the left equation in~(\ref{rhovcond}),
we conclude that~$\rho_x = 1/2$. Furthermore, the Pauli vectors must all
have the same length. In view of~(\ref{rhovc2}), this means that
\[ |\vec{v}_x| \;=\; v \;\geq\; \frac{1}{2} \qquad {\mbox{for all $x \in M$}}\:. \]
Moreover, the angles between any two vectors~$\vec{v}_x, \vec{v}_y$
with~$x \neq y$ must coincide. The only configuration with these properties
is that the vectors~$\vec{v}_x$ form the vertices of a tetrahedron, see Figure~\ref{fig1}.
Labeling the vertices by the corresponding space-time points
distinguishes an orientation of the tetrahedron; in particular, the two tetrahedra
in Figure~\ref{fig1} cannot be mapped onto each other by an orientation-preserving
rotation~(\ref{Rot}). This also implies that with the transformation~(\ref{Rot}) we cannot realize
odd permutations of the space-time points. Hence the fermion system cannot be permutation
symmetric, a contradiction.

\begin{figure}[t]
\begin{center}
\includegraphics[width=4cm]{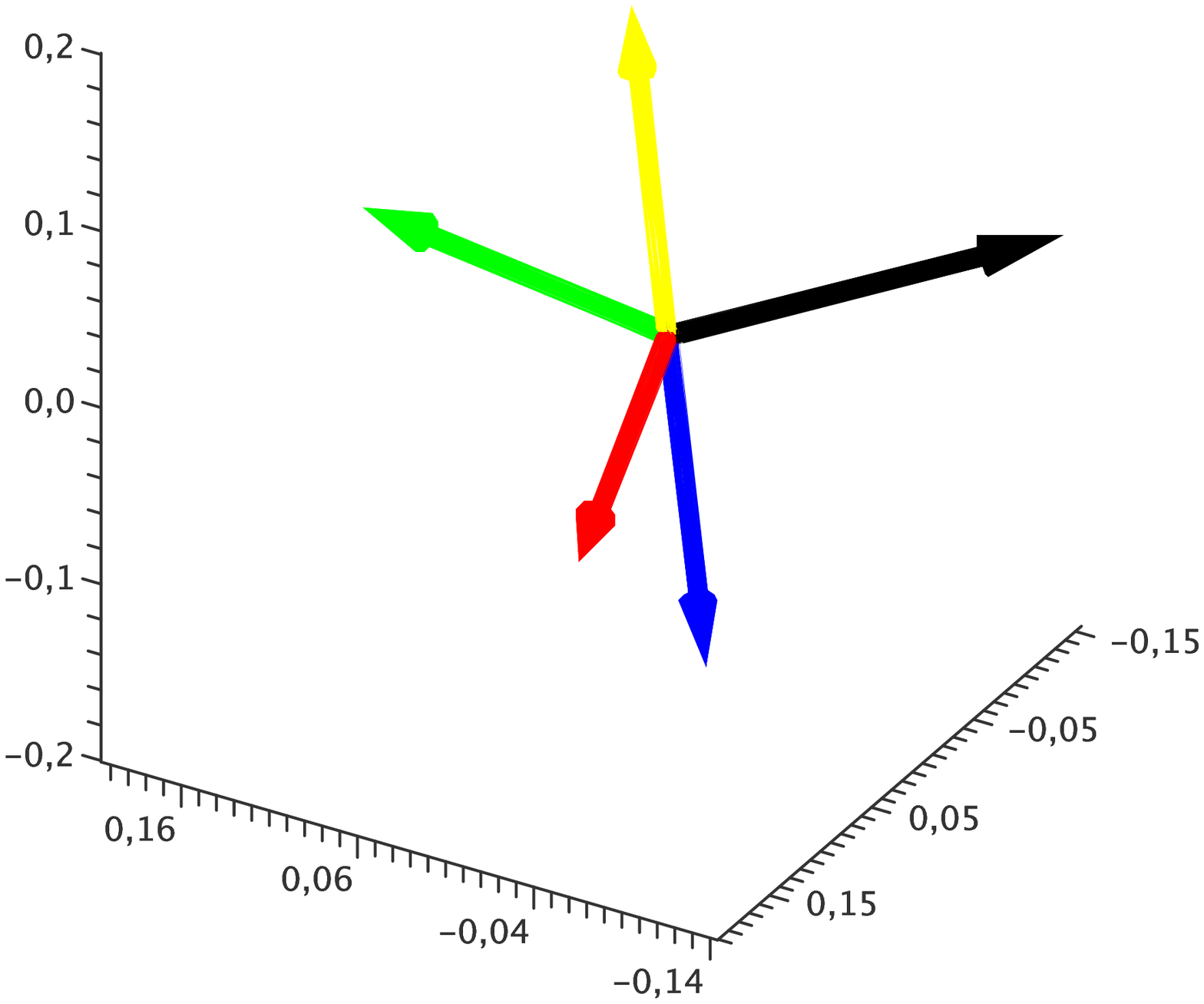}
\hspace*{-0.4cm}
\includegraphics[width=3.8cm]{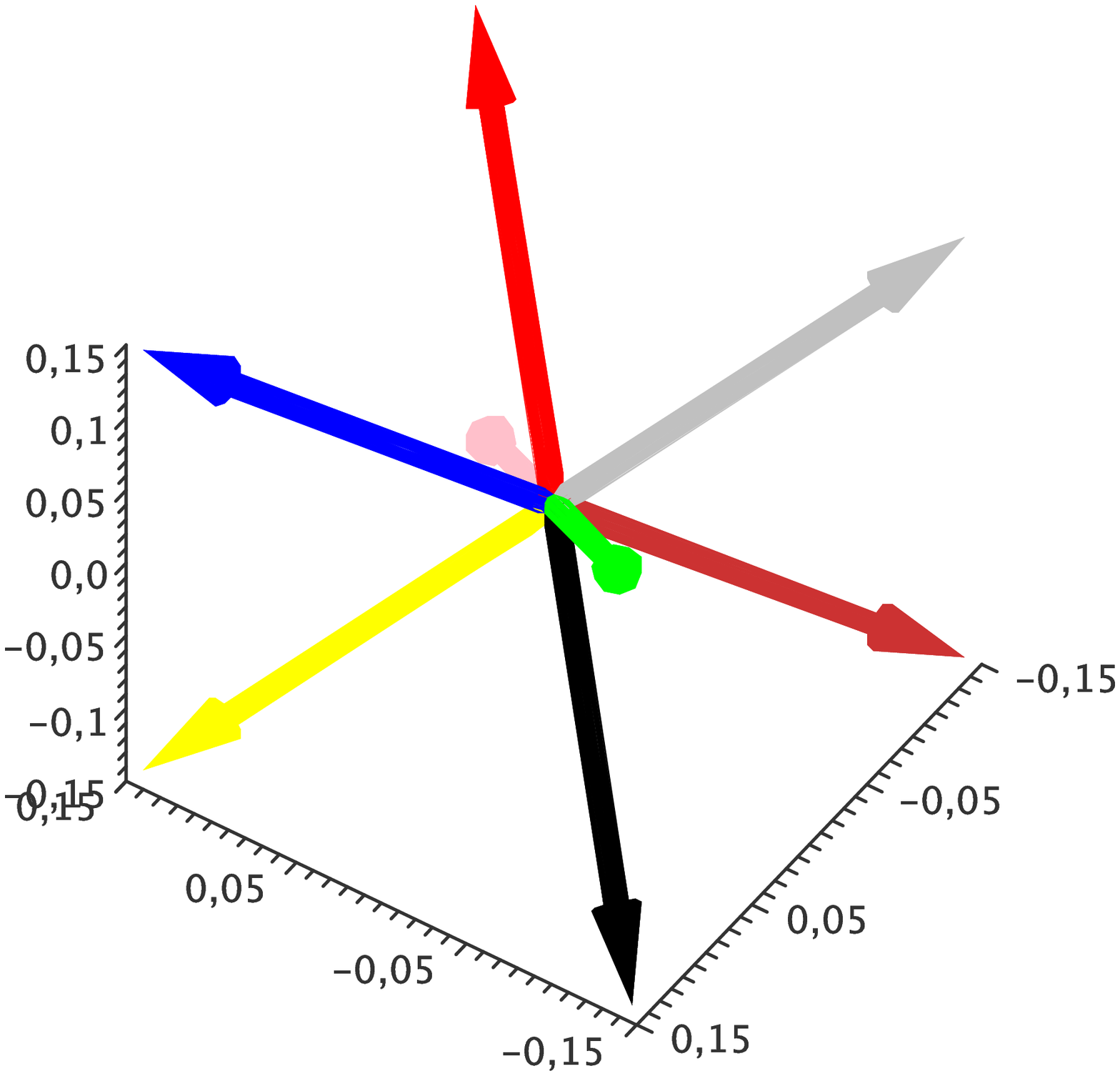}
\hspace*{-.3cm}
\includegraphics[width=4.8cm]{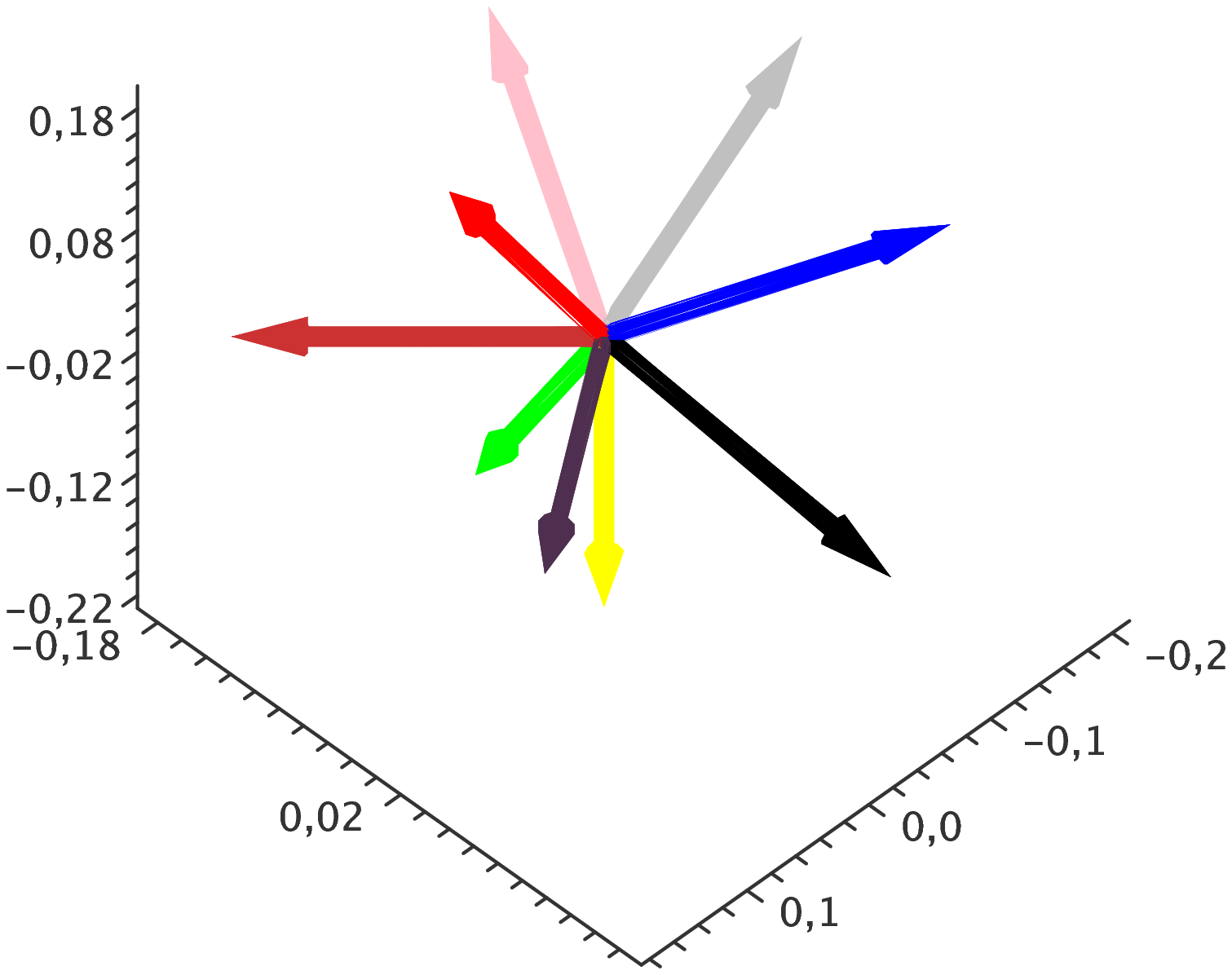}
\caption{Pauli vectors of the minimizers for five, eight and nine space-time points}
\label{fig2}
\end{center}
\end{figure}

Theorem~\ref{thmend} makes the effect of spontaneous symmetry breaking rigorous
and shows that the fermionic projector induces non-trivial relations between the
space-time points. But unfortunately, the theorem gives no information on what the resulting smaller
outer symmetry group is, nor how the induced relations on the space-time points look like.
For answering these questions, the setting of Theorem~\ref{thmend} is too general,
because the particular form of our variational principle becomes important. The basic question
is which symmetries the minimizers have.
In~\cite{DFS} the minimizers of the critical action are constructed numerically
for two particles and up to nine space-time points. For four space-time points, the Pauli
vectors of the minimizers indeed form a tetrahedron. In Figure~\ref{fig2}, the
Pauli vectors of minimizers are shown in a few examples.
Qualitatively, one sees that for many space-time points, the vectors~$\vec{v}_x$
all have approximately the same length~$2/m$ and can thus be identified with points on a
two-dimensional sphere of radius~$2/m$. The critical variational principle aims
at distributing these points uniformly on the sphere. The resulting structure is similar
to a lattice on the sphere. Thus we can say that for the critical
action in the case~$f=2$ and in the limit~$m \rightarrow \infty$, there is numerical evidence
that the spontaneous symmetry breaking leads to the emergence of
the structure of a two-dimensional lattice.

The above two-particle systems exemplify the spontaneous generation of additional
structures in discrete space-time. However, one should keep in mind that for the transition
to Minkowski space one needs to consider systems which involve many particles and are thus
much more complicated. Before explaining how this transition is supposed to work,
we need to consider how causality arises in the discrete framework.

\section{Emergence of a Discrete Causal Structure} \label{sec4}
In an indefinite inner product space, the eigenvalues of a self-adjoint operator~$A$ need
not be real, but alternatively they can form complex conjugate pairs (see~\cite{GLR}
or~\cite[Section~3]{F1}).
This simple fact can be used to introduce a notion of causality.
\begin{Def} {\bf{(discrete causal structure)}} \label{dcs}
Two discrete space-time points~$x,y \in M$ are called {\bf{timelike}}
separated if the roots~$\lambda_j$ of the characteristic polynomial
of~$A_{xy}$ are all real. They are said to be
{\bf{spacelike}} separated if all the~$\lambda_j$ form complex
conjugate pairs and all have the same absolute value.
\end{Def}
As we shall see in Section~\ref{sec5} below, for Dirac spinors in Minkowski space
this definition is consistent with the usual notion of causality.
Moreover, the definition can be understood within discrete space-time in that it reflects
the structure of the critical action. Namely, suppose that two space-time points~$x$ and~$y$
are spacelike separated. Then the critical Lagrangian~(\ref{Lcrit}) vanishes.
A short calculation shows that the first variation~${\mathcal{M}}[A_{xy}]$,
(\ref{fvar}), also vanishes, and thus~$A_{xy}$ does not enter the Euler-Lagrange
equations. This can be seen in analogy to the usual notion of causality
that points with spacelike separation cannot influence each other.

In~\cite[Section~5.2]{DFS} an explicit example is given where
the spontaneous symmetry breaking gives rise to a non-trivial discrete causal structure.
We now outline this example, omitting a few technical details.
We consider minimizers of the variational principle with constraint~(\ref{vary}, \ref{constraint})
in the case~$n=1$, $f=2$ and $m=3$.
We found numerically that in the range of~$\kappa$
under consideration here, the minimizers are permutation symmetric.
Thus in view of~(\ref{rhovcond}, \ref{rhovc2}),
the local correlation matrices are of the form~(\ref{Fxd}) with
\[ |\vec{v}_x| \;=:\; v \;\geq\;  \frac{2}{3} \;=\; \rho_x \:, \]
and the three Pauli vectors form an equilateral triangle. In~\cite[Lemma~4.4]{DFS} it is
shown that any such choice of local correlation matrices can indeed be realized by
a fermionic projector. Furthermore, it is shown that all fermionic projectors corresponding to
the same value of~$v$ are gauge equivalent. Thus we have, up to gauge transformations,
a one-parameter family of fermionic projectors, parametrized by~$v \geq 2/3$.

\begin{figure}
\begin{center}
 \includegraphics[width=11cm]{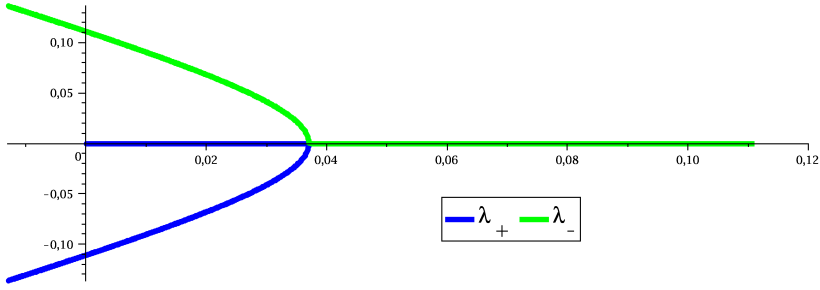}
 \caption{Plots of $\lambda_+$ and $\lambda_-$ in the complex plane for varying~$v$.}
\label{fig3}
\end{center}
\end{figure}

We again represent the fermionic projector as in~(\ref{2braket}). Then the closed
chain can be written as
\[ A_{xy} \;=\; \sum_{i,j=1}^2 E_x \,u_i \ket \bra u_i | E_y \,
u_j \ket \bra u_j | E_x\:. \]
Using the identity~$\det(BC - \lambda)=\det(CB-\lambda)$, cyclically commuting the
factors does not change the spectrum, and thus~$A_{xy}$ is isospectral to the matrix
\[ \sum_{i=1}^2 \bra u_j | E_x \, u_i \ket \bra u_i | E_y \, u_k \ket \;=\;
(F_x F_y)_{jk} \:. \]
This makes it possible to express the roots~$\lambda_\pm$ of the characteristic polynomial
of~$A_{xy}$ in terms of the local correlation matrices.
A direct computation gives~(see~\cite[Proposition~4.1]{DFS})
\[ \lambda_\pm \;=\; \frac{1}{4} \, \left( \rho_x \rho_y + \vec{v}_x \vec{v}_y \:\pm\:
\sqrt{|\rho_x \vec{v}_y + \rho_y \vec{v}_x|^2 - |\vec{v}_x \times \vec{v}_y|^2 } \right) \:. \]
If~$x=y$, the cross product vanishes, and thus the~$\lambda_\pm$ are real.
Hence each space-time point has timelike separation from itself
(we remark that this is
valid in general, see~\cite[Proposition~2.7]{DFS} and~\cite[Lemma~4.2]{F1}).
In the case~$x \neq y$, the eigenvalues $\lambda_+$ and~$\lambda_-$ are shown
in Figure~\ref{fig3} for different values of~$v$.
If~$v=2/3$, the eigenvalue $\lambda_-$ vanishes, whereas~$\lambda_+=1/9$.
If~$v=\frac{4 \sqrt{3}}{9}$, the values of~$\lambda_-$ and~$\lambda_+$ coincide.
If~$v$ is further increased, the $\lambda_\pm$ become complex and
form a complex conjugate pair. Hence different space-time points 
have timelike separation if~$v \leq \frac{4 \sqrt{3}}{9}$, whereas they have
spacelike separation if~$v > \frac{4 \sqrt{3}}{9}$. In the latter case the discrete
causal structure is non-trivial, because some pairs of points have spacelike
and other pairs timelike separation.
Finally, a direct computation of the constraint~(\ref{constraint})
gives a relation between~$v$ and~$\kappa$. One finds that $v>\frac{4 \sqrt{3}}{9}$
if and only if~$\kappa > \frac{68}{81}$. We conclude that in the case~$\kappa>\frac{68}{81}$,
the spontaneous symmetry breaking leads to the emergence of non-trivial discrete causal
structure.

We point out that the discrete causal structure of Definition~\ref{dcs}
differs from the definition of a causal set (see~\cite{causal}) in that 
it does not distinguish between future and past directed separations.
In the above example with three space-time points, the resulting discrete causal structure
is also a causal set, albeit in a rather trivial way where each point has timelike separation
only from itself.

\section{A First Connection to Minkowski Space} \label{sec5}
In this section we describe how to get a simple connection between
discrete space-time and Minkowski space. In the last Sections~\ref{sec6}--\ref{sec8}
we will proceed by explaining the first steps towards making this intuitive picture precise.
The simplest method for getting a correspondence to relativistic quantum mechanics in
Minkowski space is to replace the discrete space-time points~$M$ by the space-time
continuum~$\R^4$ and the sums over~$M$ by space-time integrals. For a
vector~$\Psi \in H$, the corresponding localization~$E_x \Psi$
should be a 4-component Dirac
wave function, and the scalar product $\bra \Psi(x) \,|\, \Phi(x) \ket$ on~$E_x(H)$
should correspond to the usual Lorentz invariant scalar product on Dirac
spinors~$\overline{\Psi} \Phi$
with~$\overline{\Psi} = \Psi^\dagger \gamma^0$ the adjoint spinor. Since this last
scalar product is indefinite of signature~$(2,2)$, we are led to choosing~$n=2$.
In view of~(\ref{diskernel}), the discrete kernel should go over to the
integral kernel of an operator~$P$ on the Dirac wave functions,
\[ (P \Psi)(x) \;=\; \int_M P(x,y)\, \Psi(y)\: d^4y \:. \]
The image of~$P$ should be spanned by the occupied fermionic states. We take Dirac's
concept literally that in the vacuum all negative-energy states are occupied by fermions
forming the so-called {\em{Dirac sea}}. Thus we are led to describe the vacuum by the
integral over the lower mass shell
\[ P(x,y) \;=\; \int \frac{d^4k}{(2 \pi)^4}\: (k \slsh+m)\:
\delta(k^2-m^2)\: \Theta(-k^0)\: e^{-ik(x-y)} \]
(here~$\Theta$ is the Heaviside function). Likewise, if we consider several
generations of particles, we take a sum of such Fourier integrals,
\beq \label{generation}
P(x,y) \;=\; \sum_{\beta=1}^g \rho_\beta \int \frac{d^4k}{(2 \pi)^4}\: (k \slsh+m_\beta)\:
\delta(k^2-m_\beta^2)\: \Theta(-k^0)\: e^{-ik(x-y)}\:,
\eeq
where~$g$ denotes the number of generations, and the~$\rho_\beta > 0$ are
weight factors for the individual Dirac seas (for a discussion of the weight
factors see~\cite[Appendix~A]{F3}).
Computing the Fourier integrals, one sees that~$P(x,y)$ is a smooth function,
except on the light cone~$\{ (y-x)^2=0\}$,
where it has poles and singular contributions (for more details see~(\ref{Pex}) below).

Let us find the connection between Definition~\ref{dcs} and the usual notion of causality
in Minkowski space. Even without computing the Fourier integral~(\ref{generation}), it is clear from the
{\em{Lorentz symmetry}} that for every~$x$ and~$y$ for which the Fourier integral
exists, $P(x,y)$ can be written as
\beq \label{Pxyrep}
P(x,y) \;=\; \alpha\, (y-x)_j \gamma^j + \beta\:\1
\eeq
with two complex coefficients~$\alpha$ and~$\beta$. Taking the complex conjugate
of~(\ref{generation}), we see that
\[ P(y,x) \;=\; \overline{\alpha}\, (y-x)_j \gamma^j + \overline{\beta}\:\1 \:. \]
As a consequence,
\beq \label{1}
A_{xy} \;=\; P(x,y)\, P(y,x) \;=\; a\, (y-x)_j \gamma^j + b\, \1
\eeq
with real parameters~$a$ and $b$ given by
\beq \label{ab}
a \;=\; \alpha \overline{\beta} + \beta \overline{\alpha} \:,\spc
b \;=\; |\alpha|^2 \,(y-x)^2 + |\beta|^2 \:,
\eeq
where~$(y-x)^2=(y-x)_j (y-x)^j$, and for the signature of the Minkowski
metric we use the convention~$(+--\,-)$.
Applying the formula~$(A_{xy} - b \1)^2 = a^2\:(y-x)^2$, one can easily compute the roots
of the characteristic polynomial of~$A_{xy}$,
\beq \label{lambdaj}
\lambda_1 \;=\; \lambda_2 \;=\; b + \sqrt{a^2\: (y-x)^2} \:,\spc
\lambda_3 \;=\; \lambda_4 \;=\; b - \sqrt{a^2\: (y-x)^2}\:.
\eeq
If the vector~$(y-x)$ is timelike, we see from the inequality~$(y-x)^2>0$ that
the~$\lambda_j$ are all real. Conversely, if the vector~$(y-x)$ is spacelike,
the term~$(y-x)^2<0$ is negative. As a consequence, the~$\lambda_j$ form complex conjugate
pairs and all have the same absolute value. This shows that for Dirac spinors in Minkowski space,
Definition~\ref{dcs} is consistent with the usual notion of causality.

We next consider the Euler-Lagrange equations corresponding to the critical
Lagrangian~(\ref{Lcrit}). If~$(y-x)$ is spacelike, the~$\lambda_j$ all have the
same absolute value, and thus the Lagrangian vanishes.
If on the other hand~$(y-x)$ is timelike, the~$\lambda_j$ as given by~(\ref{lambdaj})
are all real, and a simple computation using~(\ref{ab}) yields that $\lambda_1 \lambda_2 \geq 0$,
so that all the~$\lambda_j$ have the same sign (we remark that this is true in more generality,
see~\cite[Lemma~2.1]{F3}). Hence the Lagrangian~(\ref{Lcrit}) simplifies to
\beq \label{Lagsimple}
{\mathcal{L}}[A_{xy}] \;=\; \left\{
\begin{array}{cl} \displaystyle
\Tr(A_{xy}^2) - \frac{1}{4}\: \Tr(A_{xy})^2
& {\mbox{if~$(y-x)$ is timelike}} \\[.8em]
0 & {\mbox{if~$(y-x)$ is spacelike}}\:. \end{array} \right.
\eeq
Now we can compute the gradient~(\ref{fvar}) to obtain (for details see~\cite[Section~2.2]{F3})
\beq \label{0}
\M[A_{xy}] \;=\; \left\{ \begin{array}{cl} \displaystyle
2 A_{xy} - \frac{1}{2}\, \Tr(A_{xy})\,\1 & {\mbox{if~$(y-x)$ is timelike}}  \\[.8em]
0 & {\mbox{if~$(y-x)$ is spacelike}}\:.
\end{array} \right.
\eeq
Using~(\ref{1}), we can also write this for timelike~$(y-x)$ as
\beq \label{0a}
\M[A_{xy}] \;=\; 2 a(x,y)\: (y-x)^j \gamma_j \:.
\eeq
Furthermore, using~(\ref{ab}) we obtain that
\beq \label{Msymm}
\M[A_{xy}] \;=\; \M[A_{yx}]\:.
\eeq
Combining the relations~(\ref{Msymm}, \ref{0a}, \ref{Pxyrep}), we find that
the two summands in~(\ref{Qxydef}) coincide, and thus
\beq \label{Qxy}
Q(x,y) \;=\; \frac{1}{2} \:{\mathcal{M}}[A_{xy}]\:P(x,y)
\eeq
(we remark that the last identity holds in full generality, see~\cite[Lemma~5.2.1]{PFP}).

\begin{figure}[tb]
\begin{center}
\scalebox{0.8}
{\includegraphics{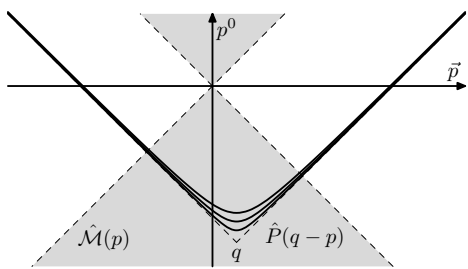}}
\caption{The convolution~${\hat{\mathcal{M}}} * \hat{P}$.}
\label{fig4}
\end{center}
\end{figure}

We point out that this calculation does not determine~$\M$ on the light cone,
and due to the singularities of~$P(x,y)$, the Lagrangian is indeed
ill-defined if~$(y-x)^2=0$. However, as an important special feature of the critical
Lagrangian, we can make sense of the Euler-Lagrange equations~(\ref{EL}),
if we only assume that~$\M$ is well-defined as a distribution.
We now explain this argument, which will be crucial for the considerations in
Sections~\ref{sec7} and~\ref{sec8}.
More precisely, we assume that the gradient of the critical Lagrangian
is a Lorentz invariant distribution, which away from the light cone coincides with~(\ref{0}),
has a vector structure~(\ref{0a}) and is symmetric~(\ref{Msymm}).
Then this distribution, which we denote for clarity by~$\tM$, can be written as
\beq \label{tM}
\tM(\xi) \;=\; 2\, \xi\slsh\: a(\xi^2)\: \Theta(\xi^2)\, \epsilon(\xi^0) \:,
\eeq
where we set~$\xi \equiv y-x$ and~$\xi\slsh \equiv \xi^j \gamma_j$, and~$\epsilon$ is the step function
(defined by $\epsilon(x)=1$ if $x \geq 0$ and~$\epsilon(x)=-1$
otherwise). We now consider the Fourier transform of the distribution~$\tM(\xi)$,
denoted by~$\hM(k)$. The factor~$\xi\slsh$ corresponds
to the differential operator~$i \Pdd_k$  in momentum space, and thus
\beq \label{odd}
\hM(k) \;=\; 2 i \,\Pdd_k \int d^4 \xi\: a(\xi^2)\: \Theta(\xi^2)\, \epsilon(\xi^0)\: e^{-i k \xi}\:.
\eeq
This Fourier integral vanishes if~$k^2<0$. Namely, due to Lorentz symmetry, in
this case we may assume that~$k$ is purely spatial, $k=(0, \vec{k})$. But then
the integrand of the time integral in~(\ref{odd}) is odd because of the step
function, and thus the whole integral vanishes.
As in~\cite{PFP}, we denote the {\em{mass cone}} as well as the upper and lower
mass cone by
\beq \label{mcdef}
{\mathcal{C}} = \{ k \:|\: k^2 > 0\}\:,\quad
{\mathcal{C}}^\vee = \{k \in {\mathcal{C}} \:|\: k^0 > 0\}\:,\quad
{\mathcal{C}}^\wedge = \{k \in {\mathcal{C}} \:|\: k^0 < 0\} \:,
\eeq
respectively. Then the above argument shows that the distribution~$\hM$ is
{\em{supported in the closed mass cone}}, ${\mbox{supp}}\, \hM \subset \overline{\mathcal{C}}$.
Next we rewrite the pointwise product in~(\ref{Qxy}) as a
convolution in momentum space,
\beq \label{ci}
\hat{Q}(q) \;=\; \frac{1}{2}\: (\hM * \hat{P})(q)
\;=\; \frac{1}{2} \int \frac{d^4p}{(2 \pi)^4}\: \hM(p)\:
\hat{P}(q-p)\:.
\eeq
If~$q$ is in the lower mass cone~${\mathcal{C}}^\wedge$, the integrand of the convolution
has compact support (see Figure~\ref{fig4}), and the integral is finite
(if however~$q \not \in \overline{\mathcal{C}^\wedge}$, the convolution
integral extends over an unbounded region and is indeed ill-defined).
We conclude that~$\hat{Q}(q)$ is well-defined inside the lower mass cone.
Since the fermionic projector~(\ref{generation}) is also supported in the lower mass
cone, this is precisely what we need in order to make
sense of the operator products~$\hat{P}(k)\,\hM(k)$ and~$\tM(k)\,\hat{P}(k)$
which appear in the commutator~(\ref{EL}). In this way we have given the
Euler-Lagrange equations a mathematical meaning.

In the above consideration we only considered the critical Lagrangian. To avoid
misunderstandings, we now briefly mention the physical significance of the
variational principle with constraint~(\ref{vary}, \ref{constraint}) and explain
the connection to the above arguments.
In order to describe a realistic physical system involving different types of fermions
including left-handed neutrinos, for the fermionic projector of the vacuum
one takes a direct sum of fermionic projectors of the form~(\ref{generation})
(for details see~\cite[\S5.1]{PFP}). On the direct summands involving the
neutrinos, the closed chain~$A_{xy}$ vanishes identically, and also the Euler-Lagrange equations
are trivially satisfied. On all the other direct summands, we want the
operator~$\M$ to be of the form~(\ref{tM}), so that the above considerations apply again.
In order to arrange this, the value of the Lagrange multiplier~$\mu$ must be larger than the
critical value~$\frac{1}{2n}$. Thus we are in the case~$\mu > \frac{1}{2n}$
where the auxiliary variational principle has no minimizers. This is why we need
to consider the variational principle with constraint~(\ref{vary}, \ref{constraint}).
Hence the fermionic projector of fundamental physics should be a minimizer of
the variational principle with constraint~(\ref{vary}, \ref{constraint}) corresponding to a
value~$\mu>\frac{1}{2n}$ of the Lagrange multiplier
(such minimizers with~$\mu>\frac{1}{2n}$ indeed exist, see~\cite[Proposition~5.2]{DFS}
for a simple example).
The physical significance of the critical variational principle
lies in the fact that restricting attention to one direct summand of the form~(\ref{generation})
(or more generally to a subsystem which does not involve chiral particles),
the Euler-Lagrange equations corresponding to~(\ref{vary}, \ref{constraint})
coincide with those for the critical Lagrangian as discussed above.
For more details we refer to~\cite[Chapter~5]{PFP}.

\section{A Static and Isotropic Lattice Model} \label{sec6}
Our concept is that for many particles and many space-time points, the mechanism explained in
Sections~\ref{sec3} and~\ref{sec4} should lead to the spontaneous emergence of the structure
of Minkowski space or a Lorentzian manifold. The transition between discrete
space-time and the space-time continuum could be made precise by proving conjectures of the following type.
\begin{Conjecture} \label{conjecture}
In spin dimension~$(2,2)$, there is a series of fermion systems in discrete space-time $(H^{(l)},  \bra .|. \ket, (E_x^{(l)})_{x \in M^{(l)}}, P^{(l)})$ with the following properties:
\begin{itemize}
\item[(1)] The fermionic projectors~$P^{(l)}$ are minimizers of the auxiliary variational
principle~(\ref{Sdef}) in the critical case $\mu=\frac{1}{4}$.
\item[(2)] The number of space-time points~$m^{(l)}$ and the number of
particles~$f^{(l)}$ scale in~$l$ as follows,
\[ m^{(l)} \;\sim\; l^4\:,\qquad f^{(l)} \;\sim\; l^3\:. \]
\item[(3)] There are positive constants~$c^{(l)}$,
embeddings~$\iota^{(l)}\::\: M^{(l)} \hookrightarrow \R^4$ and
isomorphisms~$\alpha^{(l)}\::\: H^{(l)} \rightarrow L^2(\Phi^{(l)}(M^{(l)}), \bra .|. \ket)$
(where~$\bra .|. \ket$ is the standard inner product on Dirac
spinors $\bra \Phi | \Psi \ket = \Phi^\dagger \gamma^0 \Psi$),
such that for any test wave functions~$\Psi, \Phi \in C^\infty_0(\R^4)^4$,
\begin{eqnarray*}
\lefteqn{c^{(l)}\!\!\!\!\! \sum_{x,y \in M^{(l)}} \!\!\! \Phi(\iota x)^\dagger \alpha^{(l)} 
E^{(l)}_x P^{(l)} E_y^{(l)} (\alpha^{(l)})^{-1} \Psi(\iota y) } \\
&&\spc \stackrel{l \rightarrow \infty}{\longrightarrow}
\int d^4x \int d^4y \:\Phi(x)^\dagger P(x,y)\, \Psi(y)\:,
\end{eqnarray*}
where~$P(x,y)$ is the distribution~(\ref{generation}).
\item[(4)] As~$l \rightarrow \infty$, the operators~$\M[A^{(l)}_{xy}]$ converge likewise to the
distribution~$\tM(\xi)$, (\ref{tM}).
\end{itemize}
\end{Conjecture}
\noindent
Similarly, one can formulate corresponding conjectures for systems involving several
Dirac seas, where the variational principle~(\ref{vary}) with constraint~(\ref{constraint})
should be used if chiral particles are involved. Moreover, it would be desirable to specify that
minimizers of the above form are in some sense generic. Ultimately, one would like to
prove that under suitable generic conditions, every sequence of minimizing fermion systems has
a subsequence which converges in the above weak sense to an interacting physical system
defined on a Lorentzian manifold.

Proving such conjectures is certainly difficult.
In preparation, it seems a good idea to analyze particular aspects of the problem.
One important task is to understand why discrete versions of Dirac sea
configurations~(\ref{generation}) minimize the critical action.
A possible approach is to analyze discrete fermion systems numerically.
In order to compare the results in a reasonable way to the continuum, one clearly needs
systems involving many space-time points and many particles. Unfortunately,
large discrete systems are difficult to analyze numerically.
Therefore, it seems a good idea to begin the numerical analysis with simplified systems,
which capture essential properties of the original system but are easier to
handle. In~\cite{FP} such a simplified system is proposed, where we employ
a spherically symmetric and static ansatz for the fermionic projector.
We now briefly outline the derivation of this model and discuss a few results.

For the derivation we begin in Minkowski space with a {\em{static}} and {\em{isotropic}}
system, which means that the fermionic projector~$P(x,y)$ depends only on the difference
$\xi=y-x$ and is spherically symmetric. We take the Fourier transform,
\beq \label{F2}
P(\xi) \;=\; \int \frac{d^4p}{(2 \pi)^4}\: \hat{P}(p)\: e^{i p \xi}\:,
\eeq
and take for~$\hat{P}$ an ansatz involving a {\em{vector-scalar structure}}, i.e.
\beq \label{Phv}
\hat{P}(p) \;=\; v_j(p) \,\gamma^j + \phi(p)\,\1
\eeq
with real functions~$v_j$ and~$\phi$.
Using the spherical symmetry, we can choose polar coordinates and carry out the angular
integrals in~(\ref{F2}). This leaves us with a two-dimensional Fourier integral from the 
momentum variables~$(\omega=p^0, k=|\vec{p}|)$ to the position
variables~$(t=\xi^0,r=|\vec{\xi}|)$. In order to discretize the system, we restrict
the position variables to a finite lattice~$\Lat$,
\[ (t,r) \;\in\; \Lat \;:=\;
\Big\{0, 2 \pi, \ldots,  \,2 \pi (N_t-1) \Big\} \times
\Big\{ 0, 2 \pi, \ldots,  \,2 \pi (N_r-1) \Big\}\:. \]
Here the integer parameters~$N_t$ and~$N_r$ describe the size of the lattice,
and by scaling we arranged that the lattice spacing is equal to~$2 \pi$.
Then the momentum variables are on the corresponding dual lattice~$\hat{\Lat}$,
\[ (\omega,k) \;\in\; \hat{\Lat} \;:=\;
\Big\{-(N_t-1), \ldots, -1,0 \Big\} \times
\Big\{ 1, \ldots, N_r \Big\}\:. \]
Defining the closed chain by~$A(t,r)=P(t,r) P(t,r)^*$, we can again introduce
the critical Lagrangian~(\ref{Ldef}) with~$\mu=\frac{1}{4}$.
For the action, we modify~(\ref{Sdef}) to
\[ {\mathcal{S}}[P] \;=\;
\sum_{(t,r) \in \Lat} \rho_t(t)\, \rho_r(r)\: {\mathcal{L}}(t,r)\:, \]
where the weight factors~$\rho_t$ and~$\rho_r$ 
take into account that we only consider positive~$t$ and that
a point~$(t,r)$ corresponds to many states on a sphere of radius~$r$.
When varying the action we need to take into account two constraints,
called the {\em{trace condition}} and the {\em{normalization condition}},
which take into account that the total number of particles is fixed and that the
fermionic projector should be idempotent.

\begin{figure}[tb]
\begin{center}
 \includegraphics[width=7cm]{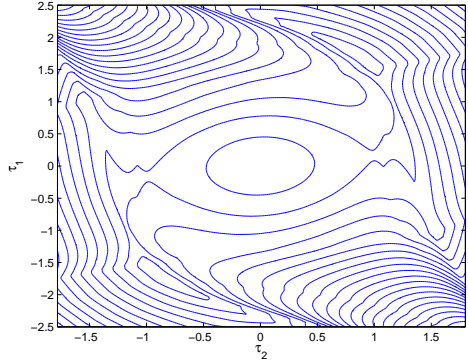}
 \caption{Action for a lattice system with two occupied states}
\label{fig5}
\end{center}
\end{figure}

In~\cite[Proposition~6.1]{FP} the existence of minimizers is proved, and we also
present first numerical results for a small lattice system. More precisely, we consider
an $8 \times 6$-lattice and occupy one state with~$k=1$ and one with~$k=2$.
The absolute minimum is attained when occupying the lattice
points~$(\omega_1=-1, k=1)$ and~$(\omega_2=-2, k=2)$. Introducing a parameter~$\tau$ by the
requirement that the spatial component of the vector~$v$ in~(\ref{Phv}) should
satisfy the relation $|\vec{v}|=\phi\, \sinh \tau$, the trace and normalization conditions
fix our system up to the free parameters~$\tau_1$ and~$\tau_2$ at the
two occupied space-time points.  In Figure~\ref{fig5} the action is shown as
a function of these two free parameters.
The minimum at the origin corresponds to the trivial
configuration where the two vectors~$v_i$ are both parallel to the $\omega$-axis.
However, this is only a local minimum, whereas
the absolute minimum of the action is attained at the two non-trivial
points~$(\tau_1 \approx 1.5, \tau_2 \approx 1)$
and~$(\tau_1 \approx -1.5, \tau_2 \approx -1)$.

Obviously, a sytem of two occupied states on an $8 \times 6$-lattice
is much too small for modelling a Dirac sea structure.
But at least, our example shows that our variational principle generates
a non-trivial structure on the lattice where the occupied points distinguish specific lattice points,
and the corresponding vectors~$v$ are not all parallel.

\section{Analysis of Regularization Tails} \label{sec7}
Another important task in making the connection to Minkowski space rigorous is
to justify the distribution~$\tM$ in~(\ref{tM}). To explain the difficulty, let us
assume that we have a family of fermion systems~$(H^{(l)},  \bra .|. \ket, (E_x^{(l)})_{x \in M^{(l)}}, P^{(l)})$ having the properties~(1)-(3) of Conjecture~\ref{conjecture}.
We can then regard the operators~$\alpha^{(l)} P^{(l)} (\alpha^{(l)})^{-1}$ as regularizations
of the continuum fermionic projector~(\ref{generation}). It is easier to consider
more generally a family of regularizations~$(P^\varepsilon)_{\varepsilon>0}$ in Minkowski
space with
\[ P^\varepsilon(x,y) \;\stackrel{\varepsilon \searrow 0}{\longrightarrow}\;
P(x,y) \quad {\mbox{in the distributional sense}}. \]
The parameter~$\varepsilon$ should be the length scale of the regularization.
In order to justify~(\ref{tM}) as well as the convolution integral~(\ref{ci}),
our regularization should have the following properties:
\begin{Def} \label{def1} The fermionic projector satisfies the assumption of
a {\bf{distributional $\M P$-product}} if the following conditions are
satisfied:
\begin{itemize}
\item[(i)] There is a distribution~$\tM(\xi)$ of the form~(\ref{tM})
such that $\lim_{\varepsilon \searrow 0} \M[A^\varepsilon_{xy}] =
\tM(\xi)$ in the distributional sense.
\item[(ii)] For every~$k$ for which~$\lim_{\varepsilon \searrow 0} \hat{Q}^\varepsilon(k)$
exists, the convolution integral~(\ref{ci}) is well-defined
and $\lim_{\varepsilon \searrow 0} \hat{Q}^\varepsilon(k) = \hat{Q}(k)$.
\end{itemize}
\end{Def}
\noindent
This notion was introduced in~\cite[\S5.6]{PFP} and used as an ad-hoc assumption on the
regularization. Justifying this assumption is not just a technicality, but seems
essential for getting a detailed understanding of how the connection between discrete space-time and
Minkowski space is supposed to work.
Namely, if one takes a simple ultraviolet regularization (for example a cutoff in momentum space),
then, due to the distributional singularity of~$P(x,y)$ on the light cone,
the product~$A^\varepsilon_{xy}=P^\varepsilon(x,y) P^\varepsilon(y,x)$ will
in the limit~$\varepsilon \searrow 0$
develop singularities on the light cone, which are ill-defined even in the distributional
sense. Thus, in order to satisfy the conditions of Definition~\ref{def1}, we need to
construct {\em{special regularizations}}, such that the divergences on the light cone
cancel. In~\cite{F3} it is shown that this can indeed be accomplished.
The method is to consider a class of spherically symmetric regularizations involving
many free parameters, and to adjust these parameters such that all the divergences
on the light cone and near the origin compensate each other. It seems miraculous that it is possible
to cancel all the divergences; this can be regarded as a confirmation for our approach.
If one believes that the regularized fermionic projector describes nature,
we get concrete hints on how the vacuum should look like on the Planck scale.
More specifically, the admissible regularizations give rise to a
{\em{multi-layer structure}} near the light cone involving {\em{several length scales}}.

In this survey article we cannot enter into the constructions of~\cite{F3}.
Instead, we describe a particular property of Dirac sea configurations which
is crucial for making the constructions work. Near the light cone, the
distribution~$P(x,y)$ has an expansion of the following form
\begin{eqnarray}
P(x,y) &=& + i C_0 \:\xi\slsh \: \frac{\mbox{PP}}{\xi^4}
+ C_1 \: \frac{\mbox{PP}}{\xi^2}
+ i C_2 \:\xi\slsh\: \frac{\mbox{PP}}{\xi^2}
+ C_3\, \log(\xi^2) + \cdots \nonumber \\
&&+ \epsilon(\xi^0) \Big( D_0 \:\xi\slsh \delta'(\xi^2)
+ i D_1 \:\delta(\xi^2) + D_2 \:\xi\slsh \delta(\xi^2)
+  i D_3 \:\Theta(\xi^2) + \cdots \Big) \qquad \label{Pex}
\end{eqnarray}
with real constants~$C_j$ and~$D_j$
(PP denotes the principal part; for more details see~\cite[Section~3]{F3}).
Let us consider the expression~$\M[A_{xy}]$, (\ref{0}), for timelike~$\xi$.
Computing the closed chain by~$A_{xy} = P(x,y)\, P(x,y)^*$, from~(\ref{Pex})
we obtain away from the light cone the expansion
\beq \label{Aexp}
A_{xy} \;=\; \frac{C_0^2}{\xi^6}+\frac{C_1^2 + 2 C_0 C_2}{\xi^4} 
+ 2 C_0 D_3 \: \frac{\xi\slsh\: \epsilon(\xi^0)}{\xi^4} \:+\: \cdots \qquad (\xi^2 > 0).
\eeq
It is remarkable that there is no contribution proportional to~$\xi\slsh/\xi^6$.
This is because the term~$\sim C_0 C_1$
is imaginary, and because the contributions corresponding to~$D_0$, $D_1$ and~$D_2$ are
supported on the light cone. Taking the trace-free part, we find
\beq \label{Msing}
\M[A_{xy}] \;=\; 4 C_0 D_3 \: \frac{\xi\slsh\, \epsilon(\xi^0)}{\xi^4} \:+\: \cdots \qquad (\xi^2 > 0).
\eeq
The important point is that, due to the specific form of the Dirac sea configuration,
the leading pole of~$\M[A_{xy}]$ on the light cone is of lower order than
expected from a naive scaling. This fact is extremely useful in the constructions of~\cite{F3}.
Namely, if we consider regularizations of the distribution~(\ref{Pex}), the
terms corresponding to~$C_0$, $C_1$ and~$C_2$ will be ``smeared out'' and will
thus no longer be supported on the light cone. In particular, the contribution~$\sim C_0 D_1$
no longer vanishes, and this {\em{vector contribution}} can be used to modify~(\ref{Msing}).
In simple terms, this effect means that the contributions by the regularization are amplified,
making it possible to modify~$\M[A^\varepsilon_{xy}]$ drastically by small
regularization terms. In~\cite{F3} we work with {\em{regularization tails}},
which are very small but spread out on a large scale~$\varepsilon^\gamma$ with~$\gamma<1$.
Taking many tails with different scales gives rise to the above-mentioned multi-layer
structure. Another important effect is that the the regularization yields {\em{bilinear
contributions}} to~$\M[A^\varepsilon_{xy}]$ of the form
$\sim i C_0 D_0 \gamma^t \gamma^r$ (with~$\gamma^r=\vec{\xi} \vec{\gamma}/|\vec{\xi}|$),
which are even more singular on the light cone than the vector contributions.
The bilinear contributions tend to make the roots~$\lambda_j$ complex
(as can be understood already from the fact that $(i \gamma^t \gamma^r)^2=-\1$).
This can be used to make a neighborhood of the light cone space-like; more precisely,
\beq \label{vanish}
\M[A^\varepsilon_{xy}] \;\equiv\; 0 \qquad
{\mbox{if~$|\xi^0| < |\vec{\xi}| + \epsilon^{\gamma}\, |\vec{\xi}|^{-\frac{1}{\alpha}}$
with~$\gamma<1$ and~$\alpha>1$}}.
\eeq

The analysis in~\cite{F3} also specifies the singularities of the distribution~$\tM$ on the
light cone (recall that by~(\ref{0}), $\tM$ is determined only away from the light cone).
We find that~$\tM$ is unique up to the contributions
\beq \label{free1}
\tM(x,y) \;\asymp\; c_0\: \xi\slsh\: \delta'(\xi^2)\: \epsilon(\xi^0) \:+\: c_1\: \xi\slsh\:
\delta(\xi^2)\: \epsilon(\xi^0)
\eeq
with two free parameters~$c_0, c_1 \in \R$. Moreover, the regularization tails give us
additional freedom to modify~$\M[A^\varepsilon_{xy}]$ near the origin~$\xi=0$.
This makes it possible to go beyond the distributional~$\M P$-product by arranging
extra contributions supported at the origin. Expressed in momentum space, we may
modify~$\hat{Q}(q)$ by a polynomial in~$Q$; namely (see~\cite[Theorem~2.4]{F3})
\begin{eqnarray}
\hat{Q}(q) \;:=\; \lim_{\varepsilon \searrow 0} \hat{Q}^\varepsilon(q)
\;=\ \frac{1}{2}\: (\hM * \hat{P})(q) + c_2 + c_3 \,q\!\!\!\slash + c_4 \,q^2 \qquad
(q \in {\mathcal{C}}^\wedge)
\label{free2}
\end{eqnarray}
with additional free parameters~$c_2, c_3, c_4 \in \R$.

\section{A Variational Principle for the Masses of the Dirac Seas} \label{sec8}
With the analysis of the regularization tails in the previous section
we have given the Euler-Lagrange equations
for a vacuum Dirac sea configuration a rigorous mathematical meaning.
All the formulas are well-defined in Minkowski space without any regularization.
The freedom to choose the regularization of the fermionic projector
is reflected by the free real parameters~$c_0, \ldots, c_4$ in~(\ref{free1}) and~(\ref{free2}).
This result is the basis for a more detailed analysis of the Euler-Lagrange equations
for vacuum Dirac sea configurations as carried out recently in~\cite{FH}. We now
outline the methods and results of this paper.

We first recall the notion of {\em{state stability}} as introduced in~\cite[\S5.6]{PFP}.
We want to analyze whether the vacuum Dirac sea configuration is a stable local
minimum of the critical variational principle within the class of static and homogeneous
fermionic projectors in Minkowski space. Thus we consider variations where we
take an occupied state of one of the Dirac seas and bring the corresponding particle
to any other unoccupied state~$q \in {\mathcal{Q}}^\wedge$. Taking into account
the vector-scalar structure in the ansatz~(\ref{generation}) and the negative definite
signature of the fermionic states, we are led to the variations (for details see~\cite[\S5.6]{PFP})
\beq \label{varstat}
\delta P = -c (k\slsh +m)\: e^{-ik(x-y)} + c (l \!\!\slash +m)\: e^{-iq(x-y)}
\eeq
with~$m \in \{m_1, \ldots,m_g\} \not\ni \sqrt{q^2}$, $k^2=l^2=m^2$ and~$k^0,l^0<0$.
We demand that such variations should not decrease the action,
\[ S[P+\delta P] \;\geq\; S[P]\quad\mbox{for all variations~(\ref{varstat})}. \]
For the proper normalization of the fermionic states, we need to consider the system
in finite $3$-volume. Since the normalization constant~$c$ in~(\ref{varstat}) tends
to zero in the infinite volume limit, we may treat~$\delta S$ as a first order perturbation.
Hence computing the variation of the action by~(\ref{dS1}), we obtain the condition
stated in the next definition.
Note that, according to~(\ref{free1}), (\ref{free2}) and~(\ref{ci}), we already know that~$\hat{Q}$
is well-defined inside the lower mass cone and has a vector scalar structure, i.e.
\beq \label{Qform}
\hat{Q} (k) \;=\; a\:\frac{k\slsh}{|k|} + b\:, \qquad k \in {\mathcal{C}}^\wedge\:,
\eeq
where we set~$|k|=\sqrt{k^2}$, and $a=a(k^2)$, $b=b(k^2)$ are two real-valued,
Lorentz invariant functions.
\begin{Def} \label{def611} \index{state stability}
The fermionic projector of the vacuum is called {\bf{state stable}}
if the functions~$a$ and~$b$ in the representation~(\ref{Qform}) of~$\hat{Q}(k)$ 
have the following properties:
\begin{itemize}
\item[(1)] $a$ is non-negative.
\item[(2)] The function $a+b$ is minimal on the mass shells,
\beq \label{ssc}
(a+b)(m^2_\alpha) \;=\; \inf_{q \in {\mathcal{C}}^\land} (a+b)(q^2)
\qquad\mbox{for all~$\alpha \in \{1,\ldots, g\}$}.
\eeq
\end{itemize}
\end{Def}

It is very helpful for the understanding and the analysis of state stability that the
condition~(\ref{ssc}) can be related to the Euler-Lagrange equations
of a corresponding variational principle. This variational principle was introduced
in~\cite{FH} for unregularized Dirac sea configurations of the form~(\ref{generation})
and can be regarded as a Lorentz invariant analog of the critical variational principle.
To define this variational principle, we expand the trace-free part of the closed chain inside
the light cone similar to~(\ref{Pex}, \ref{Aexp}) as follows,
\[ A_0(\xi) \;:=\; A_{xy} - \frac{1}{4}\: \Tr(A_{xy}) \;=\;
\xi\slsh \:\epsilon(\xi^0) \left( \frac{\m_3}{\xi^4} + \frac{\m_5}{\xi^2}
+ {\mathcal{O}}(\log \xi^2) \right) , \]
where the coefficients~$\m_3$ and~$\m_5$ are functions of the parameters~$\rho_\beta$
and~$m_\beta$ in~(\ref{generation}). Using the simplified form~(\ref{Lagsimple}) of the
critical Lagrangian, we thus obtain for~$\xi$ in the interior of the light cone the expansion
\[ {\mathcal{L}} \;=\; \Tr(A_0(\xi)^2)
\;=\; \frac{\m_3}{\xi^6} + \frac{2 \m_3 \m_5}{\xi^4} + {\mathcal{O}}(\xi^{-2}
\log \xi^2) \qquad (\xi^2>0). \]
The naive adaptation of the critical action~(\ref{Sdef}) would be to integrate~${\mathcal{L}}$
over the set~$\{ \xi^2 > 0\}$ (for details see~\cite[Section~2]{FH}). However, this integral
diverges because the hyperbolas~$\{\xi^2 = {\mbox{const}} \}$, where~${\mathcal{L}}$
is constant, have infinite measure.
To avoid this problem, we introduce the variable~$z=\xi^2$ and consider instead the
one-dimensional integral~$\int_0^\infty {\mathcal{L}}(z)\, z \,dz$, which has the
same dimension of length as the integral~$\int {\mathcal{L}}\, d^4\xi$.
Since this new integral is still divergent near~$z=0$, we subtract suitable counter terms and set
\beq \label{Liact}
{\mathcal{S}} \;=\; \lim_{\varepsilon \searrow 0} \left( \int_\varepsilon^\infty
{\mathcal{L}}(z)\: z\, dz \:-\: \frac{\m_3^2}{\varepsilon} \:+\:
2 \,\m_3\, \m_5\, \log \varepsilon \right) .
\eeq
In order to build in the free parameters~$c_0, c_1$ in~(\ref{free1}) and~$c_2, c_3$
in~(\ref{free2}), we introduce the {\em{extended action}} by adding extra terms,
\beq \label{Sext}
{\mathcal{S}}_{\mbox{\scriptsize{ext}}} \;=\; {\mathcal{S}} \:+\: F(\m_3, \m_5)
+ c_3 \sum_{\beta=1}^g \rho_\beta\, m_\beta^4 + c_4 \sum_{\beta=1}^g \rho_\beta\, m_\beta^5 \:,
\eeq
where~$F$ is an arbitrary real function (note that the parameter~$c_2$ in~(\ref{free2}) is
irrelevant for state stability because it merely changes the function~$b$ in~(\ref{Qform})
by a constant).
In our {\bf{Lorentz invariant variational principle}} we minimize~(\ref{Sext}), varying
the parameters~$\rho_\beta$ and~$m_\beta$ under the constraint
\[ \sum_{\beta=1}^g m_\beta\, \rho_\beta^3 \;=\; {\mbox{const}}\:. \]
This constraint is needed to rule out trivial minimizers; it can be understood as replacing
the condition in discrete space-time that the number of particles is fixed.

\begin{figure}[tb]
\begin{center}
{\includegraphics{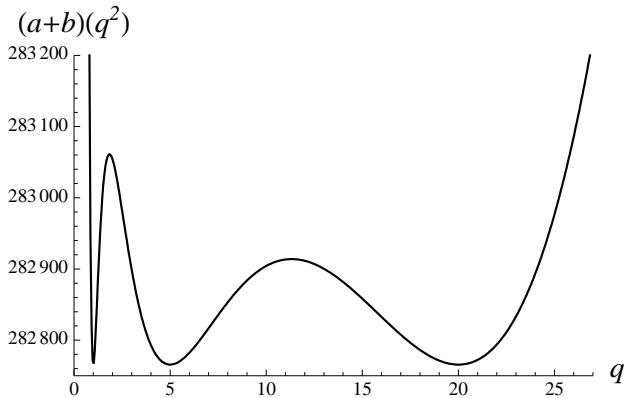}}
\caption{A state stable Dirac sea structure with three generations}
\label{fig6}
\end{center}
\end{figure}

In~\cite{FH} it is shown that, allowing for an additional ``test Dirac sea''
of mass $m_{g+1}$ and weight~$\rho_{g+1}$ (with~$\rho_{g+1}=0$
but~$\delta \rho_{g+1} \neq 0$), the corresponding Euler-Lagrange equations
coincide with~(\ref{ssc}). The difficult point in the derivation is to
take the Fourier transform of the Lorentz invariant action and to
reformulate the~$\varepsilon$-regularization in~(\ref{Liact}) in momentum space.
In~\cite{FH} we proceed by constructing numerical solutions of the Euler-Lagrange
equations which in addition satisfy the condition~(1) in Definition~\ref{def611}. We thus
obtain state stable Dirac sea configurations. Figure~\ref{fig6}
 shows an example with three generations
and corresponding values of the parameters~$m_1=1$, $m_2=5$, $m_3=20$
and~$\rho_1=1$, $\rho_2=10^{-4}$, $\rho_3=9.696 \times 10^{-6}$.

\section{The Continuum Limit} \label{sec9}
The continuum limit provides a method for analyzing the Euler-Lagrange equations (\ref{EL})
for interacting systems in Minkowski space. For details and results we refer
to~\cite[Chapters~6-8]{PFP}; here we merely put the procedure of the continuum limit in the
context of the methods outlined in Sections~\ref{sec5}--\ref{sec8}.
As explained in Section~\ref{sec7}, the regularization yields bilinear contributions
to~$A^\varepsilon_{xy}$, which make a neighborhood of the light cone spacelike~(\ref{vanish}).
Hence near the light cone, the roots~$\lambda_j$ of the characteristic polynomial
form complex conjugate pairs and all have the same absolute value,
\beq \label{absl}
|\lambda_i| = |\lambda_j| \qquad {\mbox{for all~$i,j$}},
\eeq
so that the critical Lagrangian~(\ref{Lcrit}) vanishes identically.
If we introduce an interaction (for example an additional Dirac wave function or
a classical gauge field), the corresponding perturbation of the fermionic projector
will violate~(\ref{absl}). We thus obtain corresponding contributions
to~$\M[A^\varepsilon_{xy}]$ in a strip of size~$\sim \varepsilon$ around the light cone.
These contributions diverge if the regularization is removed. For small~$\varepsilon$,
they are much larger than the contributions by the regularization tails as discussed in
Section~\ref{sec7}; this can be understood from the fact that they are much closer to the
light cone.
The formalism of the continuum limit is obtained by an expansion of these divergent
contributions in powers of the regularization length~$\varepsilon$.
The dependence of the expansion coefficients on the regularization is analyzed
using the {\em{method of variable regularization}}; we find that this dependence
can be described by a small number of free parameters, which take into account
the unknown structure of space-time on the Planck scale.
The dependence on the gauge fields can be analyzed explicitly using the
method of integration along characteristics or, more systematically, by performing a
{\em{light-cone expansion}} of the fermionic projector. In this way, one can
relate the Euler-Lagrange equations to an effective interaction
 in the framework of second quantized Dirac fields and
classical bosonic fields.

\section{Outlook and Open Problems}
In this paper we gave a detailed picture of the transition from discrete space-time to
the usual space-time continuum. Certain aspects have already been worked out
rigorously. But clearly many questions are still open.
Generally speaking, the main task for making the connection between discrete
space-time and Minkowski space rigorous is to clarify the symmetries and the
discrete causal structure of the minimizers for discrete systems involving many particles and many
space-time points. More specifically, we see the following directions for future work:
\begin{enumerate}
\item {\em{Numerics for large lattice models:}} The most direct method to
clarify the connection between discrete and continuous models is to the static
and isotropic lattice model~\cite{FP} for systems which are so large that they can
be compared in a reasonable way to the continuum. Important questions are whether
the minimizers correspond to
Dirac sea configurations and what the resulting discrete causal structure is. The next step
will be to analyze the connection to the regularization effects described in~\cite{F3}.
In particular, does the lattice model give rise to a multi-layer structure near the light cone?
What are the resulting values of the constants~$c_0, \ldots, c_4$ in~(\ref{free1}, \ref{free2})?
\item {\em{Numerics for fermion systems in discrete space-time:}} For more than two particles, almost nothing is known about the minimizers of our variational principles. 
A systematic numerical study could answer the question whether for many particles
and many space-time points, the minimizers have outer symmetries which
can be associated to an underlying lattice structure. 
A numerical analysis of fermion systems in discrete space-time
could also justify the spherically symmetric and static ansatz in~\cite{FP}.
\item {\em{Analysis and estimates for discrete systems:}} In the critical case,
the general existence problem for minimizers is still open. Furthermore, using
methods of~\cite{F2}, one can study fermion systems with prescribed outer symmetry
analytically. One question of interest is whether for minimizers the discrete causal
structure is compatible with the structure of a corresponding causal set.
 It would be extremely useful to have a method for analyzing the minimizers
asymptotically for a large number of space-time points and many particles.
As a first step, a good approximation technique (maybe using methods from quantum
statistics?) would be very helpful.
\item {\em{Analysis of the Lorentz invariant variational principle:}}
In~\cite{FH} the variational principle for the masses of the vacuum Dirac seas is introduced and
analyzed. However, the existence theory has not yet been developed.
Furthermore, the structure of the minimizers still needs to be worked out systematically.
\item {\em{Analysis of the continuum limit:}}
It is a major task to analyze the continuum limit in more detail.
The next steps are the derivation of the field equations and the
analysis of the spontaneous symmetry breaking of the chiral gauge
group. Furthermore, except for~\cite[Appendix~B]{F4}, no
calculations for gravitational fields have been made so far.
The analysis of the continuum limit should also give constraints for the weight
factors~$\rho_\beta$ in~(\ref{generation}) (see~\cite[Appendix~A]{FH}).
\item {\em{Field quantization:}}
As explained in~\cite[\S3.6]{PFP}, the field quantization effects
should be a consequence of a ``discreteness'' of the interaction described by our
variational principle. This effect could be studied and made precise for small discrete systems.
\end{enumerate}
Apart from being a challenge for mathematics, these problems have the physical
perspective of clarifying the microscopic structure
of our universe and explaining the emergence of space and time. \\[1em]

\end{document}